\documentstyle[11pt,paspconf,epsf]{article}

\begin{document}

\title{Magnetorotational Supernova Explosion -
       2D Numerical Simulation}

\author{N. V. Ardeljan}
\affil{Department of Computational Mathematics and Cybernetics,
       Moscow State University, Vorobjevy Gory, Moscow B-234 119899,
       Russia, ardel@redsun.cs.msu.su}

\author{ G. S. Bisnovatyi-Kogan, S. G. Moiseenko}
\affil{Space Research Institute, Profsoyuznaya 84/32,
       Moscow 117810, Russia,
       gkogan@mx.iki.rssi.ru, moiseenko@mx.iki.rssi.ru}

\begin{abstract}

Results of 2D numerical simulation of magnetorotational
mechanism of supernova explosion are presented. It is shown
that due to the differential rotation of the star toroidal
component of magnetic field appears and grows with time.
Angular momentum transfers outwards by the toroidal magnetic
field. With the evolution of the process part of the envelope of
the star is throwing away. The amount of thrown away mass and
energy are estimated. The results of the simulation are
qualitatively correspond to supernova explosion picture.

\end{abstract}


\section{Introduction}

One of the possible mechanisms of supernova explosions is
magnetorotational mechanism, suggested by Bisnovatyi-Kogan
(1970). It is based on the transformation of the gravitational
energy to the energy of explosion by magnetic field.
Poloidal magnetic field in differentially rotating star is
twisted and toroidal component of the magnetic field appears and
grows with the time. When the force produced by the magnetic
field becomes comparable to the gravitational force it pushes
the matter of the star outward.

A numerical simulation of the magnetorotaional supernova
explosion in 1D approximation has been made by Bisnovatyi-Kogan
{\it et al.} (1976), Ardeljan {\it et al.} (1979),
M\"uller and Hillebrandt (1979).

Pioneer 2D simulations of supernova explosion with magnetic
field has been made by Le Blank and Wilson (1970).They used
unrealistically large magnetic fields: initial magnetic energy
was comparable to the gravitational energy of the star. More
realistic case was considered by Ohnishi (1985). As result of
the simulations they got axial jets.  Simulation of the collapse
of magnetized rotating gas cloud in 2D has been done in number
of papers, see for reference Ardeljan {\it et al.} (1996b).

Realistic magnetic fields in the stars are rather small (the
ratio between magnetic and gravitational energies is about
$10^{-6} - 10^{-8}$. Such a small magnetic fields are the main
difficulty for the numerical simulations, because of very
small hydrodynamic time scale and huge time scale of the magnetic
field amplification.

Numerical simulation of such stiff problem requires application
of the implicit numerical methods, which are free from Courant
restriction on the time step.

\section{Mathematical formulation of the problem, numerical method}

The problem of the collapse of rotating magnetized gas cloud
is described by the set of magnetohydrodynamical equations with
selfgravitation for the gas with infinite conductivity
(Landay, Lifshitz (1984)).  Axial
($\frac {\partial} {\partial \varphi}=0$) symmetry
and symmetry to the equatorial plane are assu\-med.
Here $\frac {\rm d} {{\rm d} t} = \frac {\partial} { \partial t} +
{\bf u} \cdot \nabla$ , ${\bf x}=(r,\varphi ,z)$ ,
${\bf u}$ - velocity vector, $\rho$ - density, $p$ - pressure,
${\bf H}$ - magnetic field vector, $\Phi$ - gravitational potential,
 $\varepsilon$ - internal energy,
$G$ - gravitational constant, $\cal R$ - universal gas constant,
$\gamma$ - adiabatic index, ${\bf H} \otimes {\bf H}$ --- tensor of
the range 2.
The problem was solved in the restricted domain. Outside the
domain a density was taken close to zero.

For the initial conditions, we assume that the cloud is rigidly
rotating uniform gas sphere with the following parameters:

$$r=3.81 \cdot 10^{16}{\rm cm}, \> \rho=1.492 \cdot 10^{-17}
{\rm g/cm}^3,$$ $$M=1.73M_\odot=3.457\cdot 10^{33} {\rm g},\>
\gamma= 5/3,\> u^r=u^z=0,$$ $$E_{\rm rot0}/\vert E_{\rm
     gr0}\vert =0.04,\quad E_{\rm in0}/\vert E_{\rm gr0}\vert
=0.01,\quad E_{\rm mag1}/\vert E_{\rm in1}\vert = 0.05.$$

The nondimensional and scaling values were  chosen as:
$$r={\tilde r}x_0, \> z={\tilde z}x_0, \> u={\tilde u}u_0,$$
$$p={\tilde p}p_0, \> \varepsilon={\tilde \varepsilon}
\varepsilon_0, \> T={\tilde T}T_0, \> \Phi={\tilde \Phi}\Phi_0,$$
$$u_0=\sqrt{4\pi G \rho_0
x_0^2}, \> t_0=\frac {x_0} {u_0},\>
p_0=\rho_0 u_0^2=\rho_0
x_0^2 t_0^{-2}, \>  T_0= \frac {u_0^2}{\cal R},$$
$$\varepsilon_0=u_0^2=x_0^2t_0^{-2}, \>
H_0=\sqrt {p_0}=x_0t_0^{-1} \rho_0^{1/2}, \>
\Phi_0=4\pi G \rho_0 x_0^2.$$
Here the values with the index zero are the scale factors and
the functions under sign "tilde"  are dimensionless functions.
All plots and figures presented below are in  nondimensional form.

Boundary values of the gravitational potential $\Phi$ have been
calculated by the formula of volume potential
(Courant\& Hilbert, 1962).
At $r=0$ we assume:
$$v_r=v_\varphi =H_r=H_\varphi=
{\rm curl}_r {\bf H}={\rm curl}_\varphi {\bf H}=0$$

and at $z=0$ we assume:
$$v_z=H_z=\frac {\partial {\rm curl}_z {\bf H}}{\partial z}=0.$$


For our simulations we used implicit conservative Lagrangian
operator difference scheme on triangular grid with
reconstruction (Ardeljan {\it et al.}, 1996a). The theory of
this numerical technique (investigation of its approximation,
stability and convergency) has been developed by Ardeljan {\it
et al.} (1987).

Grid restructuring procedure allows us to overcome grid
overlapping situation in gas flows with nonuniform contractions
or expansions or flows with vortexes. Due to the implicitness of
the scheme it was possible to time steps, which are much larger
then Courant time step.

The details of the scheme used in this paper will be published
elsewhere.

\section {Results}

Initial number of grid knots was 5000. At the initial time moment
($t=0$) magnetic field is "switched off", because it does not
influence significantly on the hydrodynamical
collapse stage, rather short in comparison with time of
the amplification of the magnetic field.

\begin{figure}
\vspace{2in}
\caption{Triangular grid and density distribution at $t=5.153$.}
\label{fig-1}
\end{figure}

After few consequent contractions and expansions the cloud comes
to the differentially rotating stationary state, at $t_1=5.153$
and consists of rapidly rotating dense core
and extended light envelope, (Figure~\ref{fig-1}).
The angular velocity of the cloud at this time presents a
function with maximum of gradient at the transition zone between
core and envelope (Figure~\ref{fig-2}).

\begin{figure}
\vspace{2in}
\caption{Distribution of the angular velocity along r-axis and
inital magnetic field at $t=5.153$.}
\label{fig-2}
\end{figure}

At the  moment $t_1=5.153$ initial magnetic field of the
following configuration (Ardeljan et al. 1996b)
(Figure.~\ref{fig-3}):

$$H_{r0}=F_r(0.5r,0.5z-2.5)-F_r(0.5r,0.5z+2.5),
H_{\varphi0}=0, $$
$$H_{z0}=F_z(0.5r,0.5z-2.5)-F_z(0.5r,0.5z+2.5),  $$
$$F_r(r,z)=k \left( \frac{2rz} {(z^2+1)^3}-
\frac{2r^3z} {(z^2+1)^5}\right),
F_z(r,z)=k\left( \frac{1}{(z^2+1)^2} -\frac
{r^2} {(z^2+1)^4} \right),$$

\noindent
has been "switched on" with coefficient $k=0.43$.
This magnetic field is symmetrical about the equatorial plane, it
has ${\rm div} {\bf H}_0 =0$, but it is not
force-free.
The choice of initial magnetic field configuration was connected
with an attempt to avoid a singularity, which could create big
numerical problems.
After the moment of "switching on" the magnetic
field, the matter starts to move to the periphery of
the envelope due to the magnetic force produced by the initial
configuration of the magnetic field, what leads to
appearance of a "finger" which grows at the outer boundary of
the cloud.  This artificial "finger" has no significant
influence on the evolution of the inner parts of the cloud.
Due to quadrupole type of symmetry of the poloidal magnetic field
the appearing toroidal component has 2 extremums.
First is situated at the equatorial plane in the zone between
core and envelope,and the second is situated above the
equatorial plane close to the $z$-axis (Figure~\ref{fig-3}).

\begin{figure}
\vspace{2in}
\caption{Toroidal magnetic field $H_\varphi$ and velocity field
at $t=10.234$.}
\label{fig-3}
\end{figure}

At $t=10.234$ toroidal part of magnetic energy reaches its
maximum. While the total energy of the toroidal component of the
magnetic field of the cloud is much smaller than its
gravitational energy, the density of the toroidal magnetic
energy in the regions near extremums of the $H_\varphi$ becomes
comparable to the density of the internal energy what
pushes out the matter of the cloud.
Starting from $t=11.342$ the kinetic energy of the pushed part
of the envelope near equatorial plane becomes higher than its
potential energy and can be ejected from the cloud.
The ejection along equatorial plane leads to
change of the shape of the cloud, which is
stretched along $r$-axis (in $r,z$ coordinates),see
(Figure~\ref{fig-4}), related to the last computational moment
$t=32.634$.

\begin{figure}
\vspace{2in}
\caption{Velocity field at $t=32.624$.}
\label{fig-4}
\end{figure}

At Figure~\ref{fig-5} time evolution
of the ejected mass of the envelope in \% to the total mass of
the cloud is presented. The ejected part of the matter contains
about 0.09\% of the gravitational energy of the stationary
cloud after collapse.

\begin{figure}
\vspace{2in}
\caption{Time evolution of the ejected mass in \% to the
total mass of the cloud.}
\label{fig-5}
\end{figure}

The problem of the collapse of the rotating magnetized gas
cloud, described above can not be considered as complete and
final explanation of the mechanism of the supernova explosions.
However the results of such simplified
formulation give evidence in favour of the
magnetorotational mechanism can as an explanation for
the problem of the supernova explosions.

Our recent simulations of the collapse of the star with  a real
equations of state and neutrino loses as in the paper by
Ardeljan {\it et al.}, 1987a show that the star contracts much
stronger and rotates much more differentially, than it was for
the simple equation of state used in this paper.
We may expect a more efficient magnetorotational explosion.

\acknowledgments

S.G.M. and G.S.B.-K. are grateful to the partial support by
the NSF grant AST-9320068, CRDF award \# Rp1-173, and RFBR
grants 96-02-16553, 97-02-26559. G.S.B.-K. thanks INTAS
grant No.93-93. N.V.A.  is thankful to the RFBR for the support
by the grant 96-01-00838.  S.G.M. is especially thankful to the
Organizing Committee for the hospitality and opportunity to
visit the workshop.

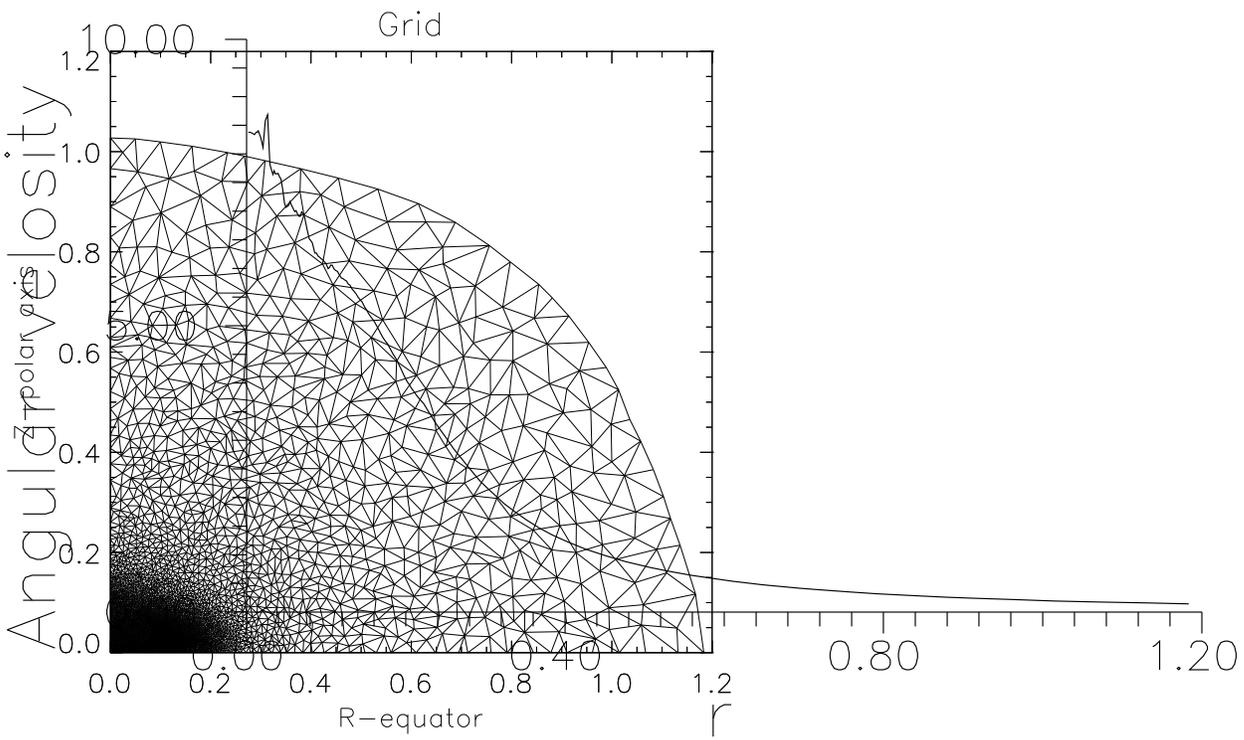
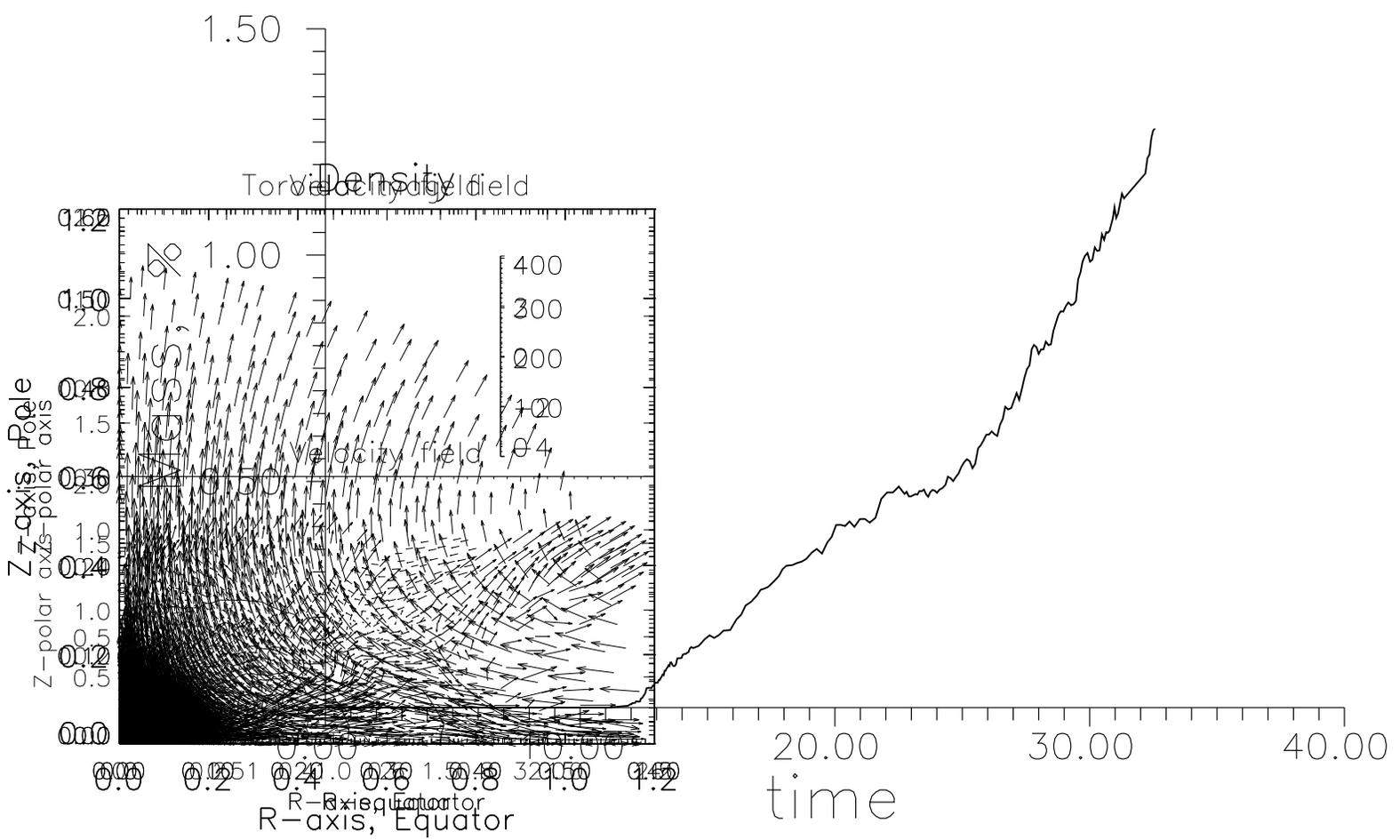
\end{document}